\documentclass[physrev,reprint,bibnotes]{revtex4-2}  

\usepackage[T1]{fontenc}
\usepackage[australian]{babel}
\usepackage[a4paper,centering,hmargin=1.75cm,vmargin=2cm]{geometry} 
\usepackage{amsmath,amssymb,graphicx,bm,microtype} 
\usepackage[dvipsnames]{xcolor} 
\usepackage{booktabs}
\usepackage[colorlinks,allcolors=blue!50!black]{hyperref}
\usepackage[all]{hypcap} 
\usepackage{cleveref} 
\usepackage{tikz}
\usetikzlibrary{quantikz2}
\usepackage{braket} 
\usepackage{upgreek,textgreek} 
\usepackage{siunitx,textcomp} 
\newcommand{\avg}[1]{\langle #1 \rangle}

\let\Im\relax \DeclareMathOperator{\Im}{Im}
\renewcommand{\d}{\mathop{}\!d}
\renewcommand{\vec}[1]{\bm{\mathrm{#1}}}

\begin{document}

\title{Any type of spectroscopy can be efficiently simulated on a quantum computer}

\author{Liam P. Flew}
\author{Ivan Kassal}
\email[Email: ]{ivan.kassal@sydney.edu.au}
\affiliation{School of Chemistry and Sydney Nano Institute, University of Sydney, NSW 2006, Australia}

\begin{abstract}
Spectroscopy is the most important method for probing the structure of molecules. However, predicting molecular spectra on classical computers is computationally expensive, with the most accurate methods having a cost that grows exponentially with molecule size. Quantum computers have been shown to simulate simple types of optical spectroscopy efficiently---with a cost polynomial in molecule size---using methods such as time-dependent simulations of photoinduced wavepackets. Here, we show that \emph{any} type of spectroscopy can be efficiently simulated on a quantum computer using a time-domain approach, including spectroscopies of any order, any frequency range, and involving both electric and magnetic transitions. Our method works by computing any spectroscopic correlation function based on the corresponding double-sided Feynman diagram, the canonical description of spectroscopic interactions. The approach can be used to simulate spectroscopy of both closed and open molecular systems using both digital and analog quantum computers.
\end{abstract}

\maketitle

The simulation of quantum systems was the first proposed application of quantum computers~\cite{Feynman1982, Lloyd1996}. Since then, quantum algorithms have been developed~\cite{AspuruGuzik2005, Kassal2008, Reiher2017, Cao2019, McArdle2020, MacDonell2021, Su2021, Babbush2023, Richerme2023, Young2023, DaJornada2025}, and experimentally demonstrated~\cite{Kassal2010, Kandala2017, Google2020, Huggins2022, MacDonell2023, Valahu2023, Richerme2023, Wang2023, Navickas2025}, for many problems in quantum chemistry, promising the solution of previously intractable computational problems.

A promising application for quantum computing is predicting spectra; spectroscopy is the most important experimental probe of molecular structure and being able to predict and interpret spectra is critical to understanding chemical behaviour. Spectroscopic techniques can be characterized by their frequencies and the order of light-matter interactions involved. Spectroscopy encompasses the entire electromagnetic spectrum, from gamma-ray M\"{o}ssbauer spectroscopy to radio-wave nuclear magnetic resonance (NMR) spectroscopy. Within this range, X-ray, UV-visible, infrared, and microwave spectroscopies probe different molecular degrees of freedom. The simplest, linear, spectroscopies measure light absorption, while nonlinear spectroscopies such as Raman, pump-probe, and 2D spectroscopies involve higher-order interactions, probing chemical dynamics. Finally, differential spectroscopy probes differences in response between two spectra. For example, dichroism spectroscopy measures the differential absorption of two light polarizations to examine anisotropic or chiral molecules.

However, accurate prediction of spectroscopic signals is computationally challenging. This challenge takes different forms depending on whether the computations are carried out in the frequency domain or in the time domain. In the frequency domain, spectra are generated by calculating eigenstates and eigenvalues to give individual spectral peaks. This approach is limited to small molecules and certain spectroscopies; in particular, when electronic and nuclear degrees of freedom are coupled (e.g., in vibronic spectroscopies), the number of relevant transitions grows exponentially with molecular size~\cite{Gordon1965, Cederbaum1976, Heller1978, Heller1981, Heller1982}. The alternative, which avoids this scaling problem, is simulating spectroscopy in the time domain, where the computational cost is determined not by the number of eigenstates but by the spectral resolution desired~\cite{Gordon1965, Cederbaum1976, Heller1978, Heller1981, Heller1982}. The time-domain approach describes spectra as Fourier transforms of correlation functions, changing spectroscopy from an eigenvalue problem to a dynamics problem. However, this approach also quickly becomes computationally expensive because it requires simulating quantum molecular dynamics, a task for which state-of-the-art methods such as multi-configuration time-dependent Hartree have costs that can grow exponentially with molecule size~\cite{Domcke2004, Worth2008}. Therefore, there remain spectroscopic problems that are computationally intractable using either approach.

Quantum computers could simulate spectroscopy efficiently (in polynomial time), but algorithms for doing so have been developed only for certain types of spectroscopy. As in classical computing, the earliest computational spectroscopy on quantum computers was in the frequency domain. In particular, many algorithms have been developed to simulate purely electronic spectroscopy, whether linear~\cite{Lee2021, Huang2022, VonBuchwald2024} or higher-order~\cite{Kharazi2025, Loaiza2024}. Frequency-domain algorithms have also been developed to simulate vibrational and vibronic spectroscopies~\cite{Huh2015, Huh2017, Shen2018, Sawaya2019, Wang2020, Jnane2021}. However, quantum frequency-domain calculations suffer from the same poor scaling as their classical counterparts because the number of transitions to be calculated scales exponentially with system size. However, because dynamics can be simulated efficiently on quantum computers, the time-domain approach can be translated into efficient quantum spectroscopy algorithms, as has been done for linear~\cite{MacDonell2023} and 2D~\cite{Bruschi2024, Gallina2025, Guimaraes2025} electronic and vibronic spectroscopy as well as for NMR~\cite{Seetharam2023} and X-ray spectroscopy~\cite{Fomichev2024}. Nevertheless, these advances have dealt with specific types of spectroscopy, making it unclear whether non-optical, higher-order, or differential spectroscopies could be simulated. 

A further problem for most existing time-domain methods is that they assume that the transition-dipole-moment operator $\mu$ can be implemented on a quantum computer using a linear combination of polynomially many (in system size) unitaries. This assumption is true for electronic spectroscopies, where often a small, constant number of electronic states is relevant. However, it fails for spectroscopies involving continuous degrees of freedom, such as vibronic spectroscopies. In that case, depending on how the continuous degree of freedom is encoded on the quantum computer, known decompositions of $\mu$ can require exponentially many unitaries.

\begin{table*}[tb]
	\centering
	\begin{tabular}{lp{8.5cm}r}
		\toprule
		Order    & Example & Representative correlation function  \\
		\midrule
        1 & Linear absorption (incl. UV-vis, infrared, X-ray, M\"{o}ssbauer) & $\avg{\mu(t_1)\mu(0)\rho(0)}$   \\
         & CD and MCD spectroscopy & $\avg{m(t_1)\mu(0)\rho(0)}$    \\ \midrule
        2 & Sum/difference frequency generation & $\avg{\mu(t_2)\mu(t_1)\mu(0)\rho(0)}$     \\ \midrule
        3 & Pump-probe spectroscopy (incl. transient absorption) &$\avg{\mu(0)\mu(t_1)\mu(t_2)\mu(0)\rho(0)}$ \\
        & 2D spectroscopy (incl. electronic, infrared, and NMR) & $\avg{\mu(t_3)\mu(t_2)\mu(t_1)\mu(0)\rho(0)}$   \\
        & Raman spectroscopy (incl. CARS and CSRS) & $\avg{\mu(0)\mu(t_1)\mu(t_1)\mu(0)\rho(0)}$   \\ 
         & 2DCD spectroscopy & $\avg{m(0)\mu(t_1)\mu(t_3)\mu(t_2)\rho(0)}$   \\ \midrule
        4 & Four-wave mixing & $\avg{\mu(t_4)\mu(t_3)\mu(t_2)\mu(t_1)\mu(0)\rho(0)}$     \\ \midrule
        5 & Fifth-order Raman spectroscopy & $\avg{\mu(t_5)\mu(t_4)\mu(t_3)\mu(t_2)\mu(t_1)\mu(0)\rho(0)}$    \\ \midrule
        $n$ &  & $\avg{\mu(t_{n})\cdots\mu(t_1)\mu(0)\rho(0)}$    \\
		\bottomrule
	\end{tabular}
	\caption{
        \textbf{Molecular spectroscopies categorized by the order of the interaction involved}, which determines the correlation functions to be computed. $\mu$ denotes the electric transition-dipole moment and $m$ the magnetic one. CD:~circular dichroism; MCD:~magnetic circular dichroism; CARS:~coherent anti-Stokes Raman spectroscopy; CSRS:~coherent Stokes Raman spectroscopy; 2DCD:~two-dimensional circular dichroism.
        }
	\label{tab:spectroscopy}
\end{table*}

Here, we show that any type of spectroscopy can be efficiently simulated on a quantum computer using the time-domain representation. Our method achieves its generality by directly constructing quantum circuits based on double-sided Feynman diagrams, the canonical representations of spectroscopic interactions to any order~\cite{Mukamel1995, Hamm2011, Yuen-Zhou2014}. In addition, it avoids the problem of inefficient unitary decompositions of $\mu$ by avoiding them altogether; instead, it computes correlation functions as derivatives of circuits that use explicitly unitary operators of the form $e^{-i\mu t}$~\cite{Pedernales2014}. 

Our approach includes all previous time-domain techniques as special cases, including purely electronic and purely vibrational ones. The algorithm's efficiency stems from the time-domain approach, meaning that it inherits the polynomial-time scaling of standard quantum algorithms for simulating time evolution~\cite{Lloyd1996, Zalka1998, Kassal2008, Low2019, Babbush2018, Cao2019, McArdle2020, MacDonell2021, Su2021, DaJornada2025}. The method is equally applicable to electric-field spectroscopies and those that involve magnetic interactions, such as circular dichroism. Finally, the method can simulate spectroscopy in open systems to account for the sensitive dependence of spectra on the molecular environment. All of these features can be implemented whether the dynamics is simulated using digital or analog quantum computers.

\section{Spectroscopic correlation functions}

Time-domain computational spectroscopy is based on correlation functions that emerge from the perturbative expansion of the light-matter interaction~\cite{Mukamel1995}. All types of spectra can be written as Fourier transforms of sums of correlation functions, with examples given in \cref{tab:spectroscopy}. Light-matter interaction is described by the Hamiltonian
\begin{equation} \label{eq:hamiltonian}
    H(t) = H_0 - \vec{\upmu} \cdot \vec{E}(t) - \vec{m} \cdot \vec{B}(t),
\end{equation}
where $H_0$ is the molecular Hamiltonian, $\vec{\upmu}$ is the electric-dipole operator, $\vec{E}(t)$ is the electric field, $\vec{m}$ is the magnetic-dipole operator, and $\vec{B}(t)$ is the magnetic field. Electric transitions are almost always stronger than magnetic ones ($\vec{\upmu}\cdot \vec{E}(t) \gg \vec{m}\cdot \vec{B}(t) $), so it is common to neglect the latter. For now, we will consider only the electric perturbation, and return to spectroscopies with magnetic effects below. For simplicity we will also assume the light is linearly polarized, $\vec{\upmu}\cdot \vec{E}(t)=\mu E(t)$.

Spectroscopy in the time domain is well understood through perturbation theory, which describes a sequence of light-matter interactions~\cite{Mukamel1995, Hamm2011}. The total spectroscopic signal is proportional to the molecular polarization $P(t)$. In the perturbative expansion, the $n$th-order term in $P(t)$ is a convolution of the $n$th-order response function $R^{(n)}(t, t_n, \ldots, t_1)$ with the electric field~\cite{Mukamel1995}:
\begin{multline}
    P^{(n)}(t) = \int_{0}^{t} \d t_n \cdots \int_{0}^{t_2} \d t_1 E(t_n)\cdots E(t_1) \\
    \times R^{(n)}(t, t_n, \ldots, t_1),
    \label{eq:convolution}
\end{multline}
where the interaction-picture response function is
\begin{multline}
\label{eq:responsefunctionnestedcommutator}
    R^{(n)}(t,t_{n},\ldots,t_1) = \\ \avg{\mu(t)[\mu(t_{n}),\cdots [\mu(t_2)[\mu(t_1), \rho(0)]]\cdots]},
\end{multline}
and where we have assumed that $E(t)=0$ for $t<0$.

Because $R^{(n)}$ is composed of $n$ nested commutators, it is a sum of $2^n$ terms, $R^{(n)} = \sum_{j=1}^{2^n} R^{(n)}_j$, where each is of the form
\begin{multline} \label{eq:bothsidescorrelationfunction}
    R^{(n)}_j(t,t_{n},\ldots,t_1) = \\ \avg{\mu(t_{K_k})\cdots\mu(t_{K_1}) \rho(0) \mu(t_{B_1})\cdots \mu(t_{B_b}) },
\end{multline}
representing a molecule interacting with the field $k$ times on the ket side and $b$ times on the bra side, where $t_{K_1} < \cdots < t_{K_k} = t$, $t_{B_1} < \cdots < t_{B_b}$, i.e., the two sets of interactions times are time ordered. The total number of interactions is one more than the order of the response function, $k+b=n+1$; the final, ($n+1$)th $\mu$ always acts on the ket and occurs because the $R^{(n)}_j$ are expectation values of $\mu$ at the final time.

Each $R^{(n)}_j$ corresponds to one of the possible double-sided Feynman diagrams, which are widely used for diagrammatically cataloging spectroscopic interactions~\cite{Mukamel1995, Hamm2011}. Identifying and constructing the relevant double-sided Feynman diagrams for a given type of spectroscopy makes it clear which correlation functions contribute to the molecular response and thus the spectrum. \Cref{fig:dsfd}a gives an example double-sided Feynman diagram.

The response function $R$ suffices to calculate any spectrum, whether it is resolved in time or frequency. While time-resolved spectra are proportional to $P(t)$ (itself a function of $R$), frequency-resolved spectra are its Fourier transforms. For example, the frequency-resolved linear absorption spectrum is
\begin{equation} \label{eq:linearabsorption}
    \sigma(\omega) \propto  \int_{-\infty}^{\infty} dt\;  e^{i\omega t} P^{(1)}(t).
\end{equation}
In the simplest case, of an electric-field consisting of a delta-function pulse, $E(t_1)= E\delta(t_1)$, we get
\begin{align}
    \sigma (\omega ) &\propto \int_{-\infty}^{\infty}dt\;  e^{i\omega t} R^{(1)}(t,0) \\
    &\propto \int_{-\infty}^{\infty} dt\;  e^{i\omega t} \avg{\mu(t) [\mu(0), \rho(0)] }.
\end{align}

In multi-pulse experiments, the electric field is a sum of individual pulses, meaning that spectra can be reported in terms of time delays between pulses. For example, the signal in transient-absorption spectroscopy depends on the delay between the pump and probe pulses, which are often assumed to be delta functions in time. If Fourier transforms are taken of some of the delays, one obtains multi-dimensional frequency-resolved spectra. For example, in 2D spectroscopy, it is common to take Fourier transforms of the delays between both the first and the last pairs of pulses, which, in the limit of delta-function pulses gives~\cite{Hamm2011}
\begin{multline}
    \sigma(\omega_3, \tau_2, \omega_1) \propto \int_{0}^{\infty} \int_{0}^{\infty} d\tau_1 \d \tau_3\; e^{i\omega_1 \tau_1} e^{i\omega_3 \tau_3} \\
    \times \sum_{n=1}^8 R_n^{(3)}(\tau_3+\tau_2 + \tau_1, \tau_2+\tau_1, \tau_1, 0),
    \label{eq:fourier}
\end{multline}
where the first pulse occurs at $t_1=0$ and where $\tau_i=t_{i+1}-t_i$ is the delay between pulses $i$ and $(i+1)$.

Also expressible using correlation functions are differential spectroscopies, which measure differences between spectroscopic signals. For example, linear dichroism spectroscopy measures differences in absorption of linear polarizations of light,
\begin{align}
    \sigma_{\mathrm{LD}}(\omega) &= \sigma_{\parallel}(\omega) - \sigma_{\perp}(\omega),
\end{align}
where $\sigma_{\parallel, \perp}(\omega)$ are given by \cref{eq:linearabsorption} for incident light polarized in two perpendicular directions.

The approach above extends to correlation functions of arbitrary combinations of electric- and magnetic-dipole operators, obtained by replacing $\mu(t)$ with $m(t)$ as required. For example, linear absorption due only to magnetic-dipole transitions involves the correlation function $\avg{m(t)m(0)\rho(0)}$. Because magnetic interactions are weak relative to electric ones, the most common interactions involve mixed correlation functions of $\mu(t)$ with $m(t)$ that are linear in $m(t)$. These include circular dichroism spectroscopy and its two-dimensional variant~\cite{Liu2023,Liu2024}. Circular dichroism is the relative absorption of left- and right-circularly polarized light~\cite{Norden2010},
\begin{align}
    \sigma_{\mathrm{CD}} (\omega) &= \sigma_{\mathrm{L}}(\omega) -\sigma_{\mathrm{R}}(\omega) \\
    & \propto \int_{-\infty}^{\infty} dt \; e^{i\omega t} \nonumber \\
    & \quad \big( \avg{\mu(t)m^{*}(0)\rho(0)}_\mathrm{L} - \avg{\mu(t)m^{*}(0)\rho(0)}_\mathrm{R} \nonumber \\
    & \quad + \avg{m(t)\mu(0)\rho(0)}_\mathrm{L} - \avg{m(t)\mu(0)\rho(0)}_\mathrm{R} \big)
 \label{eq:circulardichroism}
\end{align}
For magnetic circular dichroism, the sample is subject to a static magnetic field, which can be modelled by adding a field-dependent term $H_0(\vec{B})$ to the molecular Hamiltonian. The resulting Zeeman splitting can break molecular symmetries and allow circular dichroism measurements of achiral molecules~\cite{Barron2004}.

\begin{figure}[tb]
	\centering
        \begin{quantikz}[]
            \lstick{$\ket{0}$} &&\gate{H}& \ctrl{1} & & \ctrl[open]{1} &\meter{\sigma_x, \sigma_y}\\
            \lstick{$\ket{\psi}$} &\qwbundle{n}&& \gate{B} & \gate{U(t)} & \gate{A} &
        \end{quantikz}
	\caption{
	\textbf{Hadamard test} for a two-point correlation function of unitary operators $A$ and $B$~\cite{Somma2003}. $G(t) = \bra{\psi} A(t) B(0) \ket{\psi}$ can be computed from measured expectation values on the ancilla qubit, $G(t)=\avg{\sigma_x}+i\avg{\sigma_y}$.
	}
	\label{fig:simplecorrelation}
\end{figure}
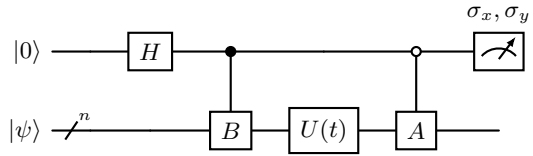

\begin{figure*}[tb] 
	\centering
	\includegraphics[width=\linewidth]{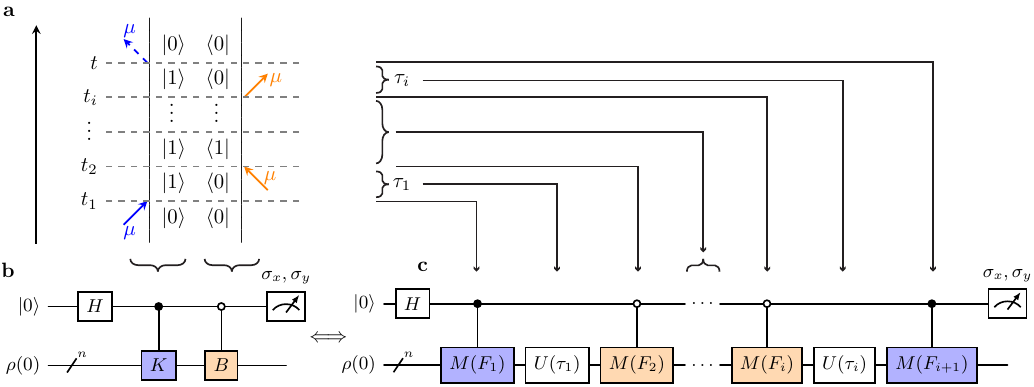}
	\caption{
	\textbf{Constructing a quantum circuit for a double-sided Feynman diagram.}
	\textbf{(a)}~Example of a double-sided Feynman diagram corresponding to a high-order correlation function. Time increases upwards and the dipole operator acts at the specified times. Additional light-matter interactions may occur between times $t_2$ and $t_i$. Absorption is shown by arrows towards either the bra or the ket and emission by arrows away from them. The dashed arrow indicates the final dipole operator, which produces the observed signal.
    \textbf{(b)}~The corresponding Hadamard test showing how the double-sided Feynman diagram maps onto a simple quantum circuit, with ket-side operators $K$ controlled off the $\ket{1}$ state of the ancilla (solid circle) and the bra-side operators $B$ controlled off $\ket{0}$ (hollow circle). $K$ and $B$ both comprise sequences of time evolution and applications of the exponentiated dipole operators, $M(F_j)=\exp(-i\mu F_j)$; here, $K=M(F_{i+1}) U(\tau_{i}) \cdots  U(\tau_1) M(F_1) $ and $B=U(\tau_{i})M(F_i) \cdots M(F_2) U(\tau_1)$.
    \textbf{(c)}~The expanded quantum circuit, with operations applied in the same order as in the Feynman diagram. As in \cref{fig:simplecorrelation}, measuring the expectation values $\avg{\sigma_x}$ and $\avg{\sigma_y}$ produces a correlation function involving $M$, which can then be differentiated to give the desired correlation function involving $\mu$.
	}
	\label{fig:dsfd}
\end{figure*}

\section{Quantum circuit}

Quantum computers can simulate time-domain spectroscopy because of their ability to efficiently simulate chemical dynamics. Here, we show how this ability can be used to calculate the response functions $R_j^{(n)}$, after which it is straightforward to classically carry out the convolutions in \cref{eq:convolution} or the Fourier transforms in, for example, \cref{eq:linearabsorption,eq:fourier,eq:circulardichroism}. Our approach to calculating $R_j^{(n)}$ is to convert any double-sided Feynman diagram into the corresponding quantum circuit. In the following examples, we give expressions for correlation functions involving $\mu$, but the algorithm works for any sequence of Hermitian observables, including cross-correlations with $m$.

Our work builds on existing quantum algorithms for calculating correlation functions. Early papers~\cite{Terhal2000, Ortiz2001, Somma2002, Somma2003} showed that two-point correlation functions
\begin{equation}
    G(t) = \bra{\psi} A(t) B(0) \ket{\psi}
\end{equation}
of unitary operators $A$ and $B$ can be calculated using the Hadamard test, a form of ancilla-qubit interferometry (see~\cref{fig:simplecorrelation})~\cite{Somma2003}. Later, the Hadamard test was generalized to $n$-point correlation functions~\cite{DelRe2024,Pedernales2014}. 

The challenge with the Hadamard test is that, in its original form, it requires the operators to be unitary. This is sometimes the case, as in spin systems where observables of interest are Pauli matrices (or products of them), which are both Hermitian and unitary. The Hadamard test could be used on non-unitary observables if they could be decomposed into linear combinations of unitaries, each of which could then be evaluated separately and the results summed. However, efficient decompositions of this kind are known in only some cases. This problem is acute in molecular spectroscopy, where correlation functions involve non-unitary observables such as $\mu$ and $m$. The exception are two-level systems, where both $\mu$ and $m$ are also unitary Pauli operators. 

To compute spectroscopic correlation functions of non-unitary operators, we use the method of~\cite{Pedernales2014}, which uses unitaries of the form $\exp(-i\mu t)$ and extracts correlation functions through differentiation. For each Hermitian operator $\mu$ in the desired correlation function, we define the unitary operator that is to be implemented on the quantum computer,
\begin{equation}
    M(F_j)=\exp \left(-i\mu F_j \right),
\end{equation}
as well as its Heisenberg-picture version, $M(t;F_j) = U^{\dagger}(t) M(F_j)U(t)$.

Spectroscopic correlation functions can then be recovered as derivatives of the correlation functions of the corresponding $M$ operators. First, $\mu$ can be obtained from a derivative of each $M(F_j)$,
\begin{equation} \label{eq:MtoMu}
    \mu=-i \frac{\partial}{\partial F_j} M(F_j) \;\biggl|_{F_j=0}.
\end{equation}
Therefore, the simplest spectroscopic correlation function is the derivative
\begin{equation}
   \bra{\psi}\mu(t)\mu(0)\ket{\psi} =(-i)^2 \frac{\partial^2}{\partial{F_1} \partial{F_2}} Q(0,t,F_1,F_2)\;\biggl|_{F_1=0, F_2=0}
\end{equation}
of the correlation function $Q(0,t,F_1,F_2)= \bra{\psi}M(t;F_2)M(0;F_1)\ket{\psi}$.
More generally, for all response functions,
\begin{multline}\label{eq:muTom}
    R_j^{(n)} (t_1,\ldots,t_n,t) =  (-i)^{n+1} \frac{\partial^{n+1}}{\partial F_1\cdots\partial F_{n+1}} \\Q_j^{(n)}(t_1,\ldots,t_n,t,F_1,\ldots,F_{n+1}),
\end{multline}
where $Q_j^{(n)}$ has same form as $R_j^{(n)}$, except that all $\mu$ operators are replaced with corresponding $M$ operators. 

To simulate spectroscopy on a quantum computer, our goal becomes constructing quantum circuits for computing $Q_j^{(n)}$, which can then be differentiated classically to obtain $R_j^{(n)}$.
Our algorithm for calculating the $Q_j^{(n)}$, illustrated in \cref{fig:dsfd}, is a Hadamard test involving operators $B$ and $K$ that group the interactions acting on the bra and the ket sides of $\rho(0)$, respectively. 
First, we convert the most generic response function (\cref{eq:bothsidescorrelationfunction}) into the corresponding $Q_j^{(n)}$,
\begin{widetext}
\begin{equation} 
    Q_j^{(n)}(t_1,\ldots,t_n,t,F_1,\ldots,F_{n+1}) = \avg{M(t_{K_k};F_{K_k})\cdots M(t_{K_1};F_{K_1}) \rho(0) M(t_{B_1};F_{B_1})\cdots M(t_{B_b};F_{B_b}) },
\end{equation}
where, as before, $k+b=n+1$. Rearranging using the cyclic property of the trace and moving to the Schrödinger picture, which corresponds to evolution on a quantum computer, this becomes
\begin{multline}
    Q_j^{(n)}(t_1,\ldots,t_n,t,F_1,\ldots,F_{n+1}) = 
    \langle \underbrace{ U^\dagger(t_{B_1})M(F_{B_1}) U(t_{B_1}) \cdots U^\dagger(t_{B_b}) M(F_{B_b}) U(t_{B_b}) U^\dagger(t_{K_k})}_{B^\dagger} \times \\
    \underbrace{ M(F_{K_k}) U(t_{K_k}) \cdots U^\dagger(t_{K_1})M(F_{K_1}) U(t_{K_1}) }_{K} \rho(0) \rangle,
    \label{eq:BandK}
\end{multline}
with $U^\dagger(t_{K_k})$ grouped into $B^{\dag}$ so that both $K$ and $B$ are time-evolved for the same total duration.

The presence of $U^\dagger$ in \cref{eq:BandK} might suggest that it is necessary to simulate reverse time evolution, which would prevent a generalization to open quantum systems. However, this is not the case because the time-evolution operators can be telescoped to forward time evolution only, using $U(t_m) U^\dagger(t_n) = U(t_m-t_n)$ if $t_m>t_n$:
\begin{multline}
    Q_j^{(n)}(t_1,\ldots,t_n,t,F_1,\ldots,F_{n+1}) =
    \langle \underbrace{ U^\dagger(t_{B_1})M(F_{B_1}) U^{\dagger}(t_{B_2}-t_{B_1}) \cdots  M(F_{B_b}) U^{\dagger}(t_{K_k}-t_{B_b})}_{B^\dagger} \times \\
    \underbrace{ M(F_{K_k}) U(t_{K_k}-t_{K_{k-1}}) \cdots U(t_{K_2}-t_{K_1})M(F_{K_1}) U(t_{K_1}) }_{K} \rho(0) \rangle .
    \label{eq:BandK_telescoped}
\end{multline}

An example of the algorithm is given in \cref{fig:dsfd} for computing a specific correlation function
\begin{align}
    Q^{(i)}_a
    &= \avg{M(t;F_{i+1})\cdots M(t_1;F_1)\rho(0)M(t_2;F_2)\cdots M(t_i;F_i)} \\
    &= \avg{M(t_2;F_2)\cdots M(t_i;F_i)M(t;F_{i+1})\cdots M(t_1;F_1)\rho(0)},
\end{align}
\end{widetext}
where the ellipses indicate that interactions may occur at intervening times. Partitioning this expression into $B$ and $K$ factors, we get
\begin{align}
    B &= \hphantom{M(F_{j+1})}U(\tau_i) M(F_i) \cdots  M(F_2) U(\tau_1), \label{eq:B} \\
    K &= M(F_{i+1}) U(\tau_i)\hphantom{M(F_j)} \cdots \hphantom{M(F_j)} U(\tau_1) M(F_1), \label{eq:K}
\end{align}
where there is only forward time evolution in both $B$ and $K$, punctuated by applications of $M(F_j)$ as required, and $\tau$ is the time interval $\tau_j=t_{j+1}-t_j$.

The quantum circuit for the Hadamard test with $B$ and $K$ is shown in \cref{fig:dsfd}b. For a system initially in pure state $\ket{\psi_0}$, the final state of the quantum circuit is
\begin{equation} \label{eq:finalstate}
    \ket{\Psi} = \frac{1}{\sqrt{2}} \bigl( \ket{0}\otimes B\ket{\psi_0} + \ket{1}\otimes K\ket{\psi_0}  \bigr).
\end{equation}
Then, as in \cref{fig:simplecorrelation}, $Q^{(i)}_a$ can be calculated as the sum of two expectation values, $\avg{\sigma_x}+i\avg{\sigma_y}$. The experiment must be repeated until the shot noise decreases below the desired error threshold. For a mixed initial state $\rho(0)$, all of the results above hold by linearity, and would allow the simulation, for example, of spectroscopy on thermal initial states. 

At this point, we have calculated $Q_j^{(n)}$ and need to numerically differentiate it to recover $R_j^{(n)}$. Classical numerical differentiation methods are well-studied~\cite{Press2007}, with the simplest being the two-point finite difference,
\begin{align}
    \mu(t) &\approx(-i)\left(\frac{M(t;\delta)-M(t;-\delta)}{2\delta} \right),
\end{align}
which has a leading-order error of $\delta^2/6\times\partial_{F_j}^3 M(t;F_j)|_{F_j=0}$~\cite{Press2007}. If two-point finite difference is not sufficiently accurate, other numerical differentiation methods can be used, such as fitting the data points to a function whose derivative can be calculated analytically~\cite{Press2007}. Numerical differentiation to compute each $\mu(t)$ requires evaluating the corresponding $M(t,F)$ at at least two values of $F$, which increases the final cost of the algorithm, which we return to in \cref{sec:cost}. Numerical differentiation can be avoided in the special case of a two-level system, where $\mu=\sigma_x$ is both Hermitian and unitary and can be used to compute the correlation functions directly. 

The full quantum circuit for our algorithm is shown in \cref{fig:dsfd}c, where time evolution of the system is punctuated by controlled applications of $M$, as required. Importantly, the time evolution itself need not be controlled because, as shown in \cref{eq:B,eq:K}, the bra and the ket components evolve together between applications of $M$. With this observation, there is a clear correspondence between the double-sided Feynman diagram in \cref{fig:dsfd}a and the resulting circuit \cref{fig:dsfd}c.

\section{Open systems}

Our algorithm can also be used to simulate spectroscopy in open quantum systems, where time evolution is non-unitary. Doing so is essential for producing spectra in a wide range of chemical and physical environments, which can significantly affect spectral linewidths, intensities, and frequencies. 

In open systems, \cref{eq:convolution,eq:responsefunctionnestedcommutator,eq:bothsidescorrelationfunction} still hold~\cite{Mukamel1995, Hamm2011}. However, in an open system, the time evolution of operators is generated by the total system-bath Hamiltonian and is no longer unitary when restricted to the system.  Therefore, in \cref{eq:convolution,eq:responsefunctionnestedcommutator,eq:bothsidescorrelationfunction}, instances of the unitary gate $U(t)$ need to be replaced with the open-system dynamics $\mathcal{U}(t)$, defined through
\begin{equation}
    \rho(t)=\mathcal{U}(t)\rho(0).
\end{equation}
It might seem that, for a second-order correlation function such as 
\begin{equation}
    \langle\mu(t)\mu(0)\rho(0)\rangle=\langle U^{\dagger}(t)\mu U(t)\rho(0)\rangle
\end{equation}
this approach leads to $\langle\mathcal{U}^{\dagger}(t)\mu\mathcal{U}(t)\rho(0)\rangle$, which is unphysical because backwards time evolution $\mathcal{U}^{\dagger}(t)$ is undefined for an open system. However, this difficulty is resolved by the formulation in \cref{eq:B,eq:K}, where the same correlation function can be simulated using only $\mathcal{U}(t)$ by making the substitution
\begin{align}
    B=U(t)\mu \quad &\to \quad B=\mathcal{U}(t)\mu \\
    K = \mu U(t) \quad &\to \quad K=\mu \mathcal{U}(t).
\end{align}
Therefore, the quantum circuits in \cref{fig:dsfd}, which implement \cref{eq:B,eq:K} also remain unchanged apart from the replacement $U\to \mathcal{U}$. 

\begin{figure}[tb]
 	\centering
 	\includegraphics[width=\columnwidth]{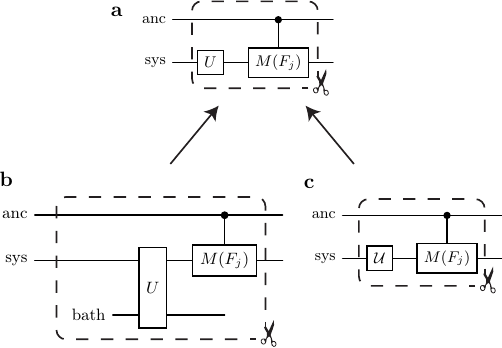}
 	\caption{
 	\textbf{Simulating open systems.} 
    \textbf{(a)}~The circuit elements of the closed-system simulation from~\cref{fig:dsfd}, which can be modified to simulate open systems.  
    \textbf{(b)}~Modification for simulating an open system with an explicit bath, where the system and the bath are governed by global unitary time evolution and the bath is traced out at the end.
    \textbf{(c)}~Modification for simulating an open system using engineered noise, where experimental noise is used to mimic the effects of open systems.
    With either modification, the circuits retain the property that the ancilla qubit only controls the application of $M$ and not the time evolution itself.
 	}
 	\label{fig:openclosed}
 \end{figure}

The specific implementation of $\mathcal{U}(t)$ depends on the model being used to describe the open-system dynamics. 
There are two ways that quantum circuits can be adapted to simulate open dynamics, whether on digital or analog quantum computers, as shown in \cref{fig:openclosed}. First (\cref{fig:openclosed}b), a bath could be explicitly represented or modelled using additional ancilla qubits~\cite{Wang2011,Gallina2025} or analog equivalents~\cite{Kim2022}. In that case, simulating the unitary system-bath evolution would result in non-unitary reduced system dynamics.
Alternatively (\cref{fig:openclosed}c), engineered noise could be injected into the quantum simulation to model dissipative effects. The latter approach is more natural in analog simulators, where protocols have been developed for injecting chemically relevant noise into simulations of vibronic dynamics~\cite{Kim2022, Olaya-Agudelo2025}.

For example, if we wished to simulate the dynamics of a system interacting with a Markovian bath, the open-system dynamics $\mathcal{U}(t)=e^{\mathcal{L}t}$ is given in terms of the Lindbladian
\begin{equation}
    \mathcal{L}=-i[H,\rho]+\sum_k \gamma_k(L_k\rho L_k^{\dagger}-\tfrac{1}{2}\{ L_k^{\dagger}L_k,\rho\}),
\end{equation}
with jump operators $L_k$ acting on the system with rates~$\gamma_k$. There are established approaches for simulating dynamics of this type on both digital~\cite{Sweke2015, Schlimgen2022, Ding2024} and analog~\cite{Olaya-Agudelo2025} quantum computers.

\section{Cost}
\label{sec:cost}
Our algorithm is efficient for any spectroscopy, with both its memory and time costs scaling polynomially with both molecular size and relevant properties in the desired spectrum.

The memory cost is equal to the number of qubits (or other computational quantum resources) required to encode the system wavefunction, plus one ancilla qubit. The cost of the wavefunction encoding depends on the particular quantum simulation algorithm being used to simulate the dynamics. For a molecule with $\eta$ atoms, this cost is generally $O(\eta)$~\cite{Lloyd1996}. This linear scaling is achieved on digital quantum computers in both first-quantized~\cite{Kassal2008, Su2021, DaJornada2025} and second-quantized~\cite{AspuruGuzik2005, McArdle2020} representations, as well as on analog quantum simulators~\cite{MacDonell2021}.

The time cost, measured in the number of required quantum gates, can be determined by multiplying the cost of each run of the circuit in \cref{fig:dsfd}c by the number of necessary repetitions, 
\begin{equation} \label{eq:Ncost}
    \mathrm{Cost} = C_U N_{\mathrm{corr}} N_{\mathrm{samples}}  N_\mathrm{shots} N_\mathrm{deriv},
\end{equation}
where the factors depend on the type of spectroscopy and the desired properties of the spectrum: $C_U$ is the cost of each run of the circuit; $N_{\mathrm{corr}}$ is the number of correlation functions $Q_j^{(n)}$ to be calculated for an $n$th order spectroscopy; $N_{\mathrm{samples}}$ is the number of points used to sample each correlation function $Q_j^{(n)}$; $N_{\mathrm{shots}}$ is the number of times the circuit must be repeated for each sampling point so that the expectation values $\avg{\sigma_x}$ and $\avg{\sigma_y}$ converge to a desired accuracy; and $N_{\mathrm{deriv}}$ is the number of different values of $F_j$ for which $Q^{(n)}_j$ needs to be calculated so that the corresponding $R^{(n)}_j$ can be determined by numerical differentiation. Each of these factors is determined by the desired properties of the final spectrum, namely its range $\omega_{\mathrm{max}}$, resolution $\Delta \omega$, and accuracy $\varepsilon$.

We estimate $C_U$ based on the fact that each run of the circuit imposes only a small overhead compared to simulating the corresponding molecular dynamics. The circuit in \cref{fig:dsfd}c is a straightforward time evolution, punctuated by applications of the controlled-$M$ operators. The additional cost of the controlled operations is minimal compared to the cost of the dynamics, and we neglect it in the following analysis. In general, the gate cost of quantum-simulation algorithms is polynomial in system size, i.e., $O(\mathrm{poly}(\eta))$. The order of this polynomial depends on the type of simulation; for example, for grid-based, first-quantized chemical dynamics, it is $O(\eta^2)$~\cite{Kassal2008, Su2021, DaJornada2025}. In terms of the scaling with the total simulation time $T$ and simulation error $\varepsilon_\mathrm{sim}$, the asymptotically best known gate cost is 
\begin{equation}
    \label{eq:CU}
    C_U\in O(T\,\mathrm{poly}(\eta)+\log\varepsilon_\mathrm{sim}^{-1}), 
\end{equation}
achieved by qubitization~\cite{Low2019, Su2021, Babbush2018}.

Naively, $N_{\mathrm{corr}}$ would equal $2\times2^n$ if the real and imaginary components were measured for each of the $2^n$ correlation functions $Q_j^{(n)}$. However, this value can be reduced by a factor of four by the symmetries of the problem. In particular, the hermiticity of the dipole moment operator implies that the response function $R_j^{(n)}$ is a sum of conjugate pairs of $Q_j^{(n)}$~\cite{Mukamel1995,Hamm2011,Kharazi2025}. Each of those pairs can therefore be evaluated using only the real or imaginary component of one of the correlation functions in the pair, as opposed to requiring both the real and imaginary components of both. For example, the first-order response function can be simplified to
\begin{align}
    R^{(1)}(t,0) &= \avg{\mu(t) [\mu(0), \rho(0)]}\\
    &=  \avg{\mu(t)\mu(0)\rho(0)} - \avg{\mu(t)\rho(0) \mu(0)} \label{eq:fourparts}\\
    &=  \avg{\mu(t)\mu(0)\rho(0)} - \avg{\mu(t)\mu(0) \rho(0)}^{\dagger}\\
    &= 2i\Im \avg{\mu(t)\mu(0)\rho(0)},
\end{align}
meaning that linear-absorption spectroscopy can be computed using a single quantity, $\Im \avg{\mu(t)\mu(0)\rho(0)}$, as opposed to the four real and imaginary parts in \cref{eq:fourparts}. This simplification reduces $N_\mathrm{corr}$ to
\begin{equation}
    N_{\mathrm{corr}}=2^{n-1}.
\end{equation}
For specific spectroscopies, $N_{\mathrm{corr}}$ may be less than this maximum because the rotating-wave approximation or phase matching can result in certain Feynman diagrams not contributing to the signal~\cite{Mukamel1995,Hamm2011}. For example, in pump-probe spectroscopy, phase matching reduces $N_{\mathrm{corr}}$ from $2^{3-1}=4$ to only 3~\cite{Mukamel1995}. 

$T$ and $N_{\mathrm{samples}}$ are governed by their Fourier counterparts $\omega_{\mathrm{max}}$ and $\Delta \omega$. To achieve a spectral resolution $\Delta\omega$, the corresponding time series must be of duration $T=2\pi/\Delta\omega$, transforming \cref{eq:CU} into
\begin{equation}
    C_U \in O(\Delta\omega^{-1} \mathrm{poly}(\eta) + \log\varepsilon_\mathrm{sim}^{-1}).
\end{equation}
Similarly, for the spectrum to reach maximum frequency $\omega_{\mathrm{max}}$, the time series must have resolution $\Delta t = \pi/\omega_{\mathrm{max}}$. This requires $T/\Delta t = 2\omega_\mathrm{\max}/\Delta\omega$ sampling points along each dimension, or a total number of samples
\begin{equation}
    N_{\mathrm{samples}}=(T/\Delta t)^n = (2\omega_\mathrm{\max}/\Delta\omega)^n.
\end{equation}

$N_{\mathrm{shots}}$ is set by the maximum error $\varepsilon_\mathrm{shot}$ in the computed spectrum. Each measurement of $\sigma_x$ and $\sigma_y$ returns 0 or 1, and needs to be repeated
\begin{equation}
    N_{\mathrm{shots}}\in O(\varepsilon_\mathrm{shot}^{-2})
\end{equation}
times to ensure the averages $\avg{\sigma_x}$ and $\avg{\sigma_y}$ are computed to within a standard error $\varepsilon_\mathrm{shot}$. This remains the standard error of every point in the frequency-domain spectrum because the Fourier transform is linear. If desired, the shot-noise scaling could be improved to the Heisenberg limit $N_{\mathrm{shots}}\in O(\varepsilon^{-1})$ using quantum amplitude estimation~\cite{Brassard2002}.

Finally, $N_{\mathrm{deriv}}$ depends on the method of numerical differentiation. The simplest case would be two-point finite difference, where 
\begin{equation}
    N_{\mathrm{deriv}}=2^{n+1},
\end{equation}
because two points are required for the derivative of $Q_j^{(n)}$ along each of the $(n+1)$ coordinates $F_j$.

Overall, using two-point finite difference for the derivatives, we arrive at the total gate cost of
\begin{equation} \label{eq:finalcost}
    \mathrm{Cost} \in O\left(\frac{2^{2n}}{\varepsilon^2 \Delta \omega} \left(\frac{2\omega_{\mathrm{max}}}{\Delta\omega}\right)^n\mathrm{poly}(\eta) \right),
\end{equation}
given maximum error $\varepsilon$, which we assume is dominated by the quadratic $\varepsilon_\mathrm{shot}$, neglecting the logarithmic $\varepsilon_\mathrm{sim}$.
This cost is polynomial in system size and, for any particular spectroscopy (i.e., a given $n$), in $\varepsilon_\mathrm{shot}$, $\omega_{\mathrm{max}}$, and $\Delta\omega$. In principle, the cost grows exponentially in $n$; however, in practice, $n$ is a small constant, rarely greater than 3 and almost never greater than 5, because high-order spectroscopy is rarely carried out. \\

\section{Discussion}

We have shown that any double-sided Feynman diagram can be converted into a quantum circuit to compute the corresponding correlation function. Therefore, because of the generality of the correlation-function approach to spectroscopy, our algorithm can simulate any type of spectroscopy, regardless of the order of the interactions, whether they are electric or magnetic, whether the system is isolated or open, or whether static external fields are included in the molecular Hamiltonian.

The efficiency of our algorithm relies on the efficiency of Hamiltonian simulation and is independent of any particular properties of the molecular Hamiltonian. Because it is possible to simulate molecular dynamics in polynomial time in system size~\cite{Lloyd1996, Zalka1998, Kassal2008, Low2019, Babbush2018, Cao2019, McArdle2020, MacDonell2021, Su2021, DaJornada2025}, the same polynomial scaling applies to spectroscopy simulation. Our approach does not constrain how the molecular system is encoded on the quantum computer or how the simulation is carried out; in particular, it is compatible with both digital~\cite{Kassal2008} and analog~\cite{MacDonell2021} molecular simulation, and whether the molecule is represented in the Born-Oppenheimer approximation or not~\cite{Kassal2008}. We neglect the cost of preparing the initial state because it is a constant overhead that would be necessary in any approach to simulating spectroscopy.

Because it uses the time-domain approach to spectroscopy, our method offers an exponential improvement in scaling over frequency-domain approaches. Its cost, given in \cref{eq:finalcost} is determined by the desired spectral properties and is polynomial in the system size $\eta$.  By contrast, frequency-domain approaches scale exponentially with $\eta$: for an $\eta$-atom molecule where each of the $3\eta-6$ vibrational modes is assumed to have $b$ accessible states, the number of spectroscopic peaks (transitions between all pairs of states) that need to be calculated in the frequency-domain approach grows exponentially as $O\left((b^{3\eta-6})^2\right)$~\cite{MacDonell2023}.

Because spectroscopy is the most direct way to probe the quantum-mechanical features of molecules, we anticipate that early applications of quantum computers in chemistry will involve the prediction and interpretation of spectroscopic signals that might be too challenging for classical computers.

\begin{acknowledgments}
We thank Ryan MacDonell, Patrick Sinnott, and Arkin Tikku for valuable discussions.
We were supported by the Australian Research Council (FT230100653), by the United States Office of Naval Research Global (N62909-24-1-2083), by the Wellcome Leap Quantum for Bio program, and by the Australian Government Research Training Program.
\end{acknowledgments}

\bibliography{bib}

\end{document}